\documentclass[journal]{IEEEtran}
\usepackage{ifpdf}
\ifpdf
\pdfoutput=1
\usepackage{graphicx,color}
\usepackage[pdftex,plainpages=false,colorlinks,citecolor=black,linkcolor=black,urlcolor=black,pdfstartview=FitH,bookmarks=true]{hyperref}
\hypersetup{pdftitle={The work of the RD51 collaboration},pdfauthor={Serge Duarte Pinto},pdfsubject={Micropattern gas detector technologies and applications}}
\else
\usepackage{graphicx}
\usepackage[dvips,plainpages=false,colorlinks,citecolor=black,linkcolor=black,urlcolor=black,pdfstartview=FitH,bookmarks=true]{hyperref}
\hypersetup{pdftitle={The work of the RD51 collaboration},pdfauthor={Serge Duarte Pinto},pdfsubject={Micropattern gas detector technologies and applications}}
\fi
\usepackage{amsmath}
\usepackage[varg]{txfonts}
\usepackage{graphicx}
\usepackage{fixltx2e}
\usepackage{cite}
\usepackage{bm}
\usepackage{amssymb}
\usepackage{footnote}
\usepackage[english]{babel}
\usepackage[tracking,spacing,kerning,letterspace=0,babel]{microtype}
\usepackage{wasysym}
\fussy
\usepackage{memhfixc}
\usepackage{booktabs}
\usepackage{multirow}
\usepackage{rotating}

\begin{document}
\title{\huge Micropattern gas detector technologies and applications\\ The work of the RD51 collaboration}
\author{\IEEEauthorblockN{Serge Duarte Pinto\IEEEauthorrefmark{1}\IEEEauthorrefmark{2}\IEEEauthorrefmark{3}, on behalf of the collaboration.}
\thanks{Manuscript received, \oldstylenums{19} November \oldstylenums{2010}.}
\thanks{\IEEEauthorblockA{\IEEEauthorrefmark{1}\textsc{Cern}, Geneva, Switzerland.}}
\thanks{\IEEEauthorblockA{\IEEEauthorrefmark{2}Physikalisches Institut der Universit\"at Bonn, Bonn, Germany.}}
\thanks{\IEEEauthorblockA{\IEEEauthorrefmark{3}Corresponding author: Serge.Duarte.Pinto@cern.ch}}}

\maketitle
\noindent
\begin{abstract}
The RD51 collaboration was founded in April 2008 to coordinate and facilitate efforts  for development of micropattern gaseous detectors (\textsc{mpgd}s).
The 75 institutes from 25 countries bundle their effort, experience and resources to develop these emerging micropattern technologies.

\textsc{Mpgd}s are already employed in several nuclear and high-energy physics experiments, medical imaging instruments and photodetection applications; many more applications are foreseen.
They outperform traditional wire chambers in terms of rate capability, time and position resolution, granularity, stability and radiation hardness.
RD51 supports efforts to make \textsc{mpgd}s also suitable for large areas, increase cost-efficiency, develop portable detectors and improve ease-of-use.

The collaboration is organized in working groups which develop detectors with new geometries, study and simulate their properties, and design optimized electronics.
Among the common supported projects are creation of test infrastructure such as beam test and irradiation facilities, and the production workshop.
\end{abstract}

\section{Introduction}
\begin{figure}
\includegraphics[width=\columnwidth]{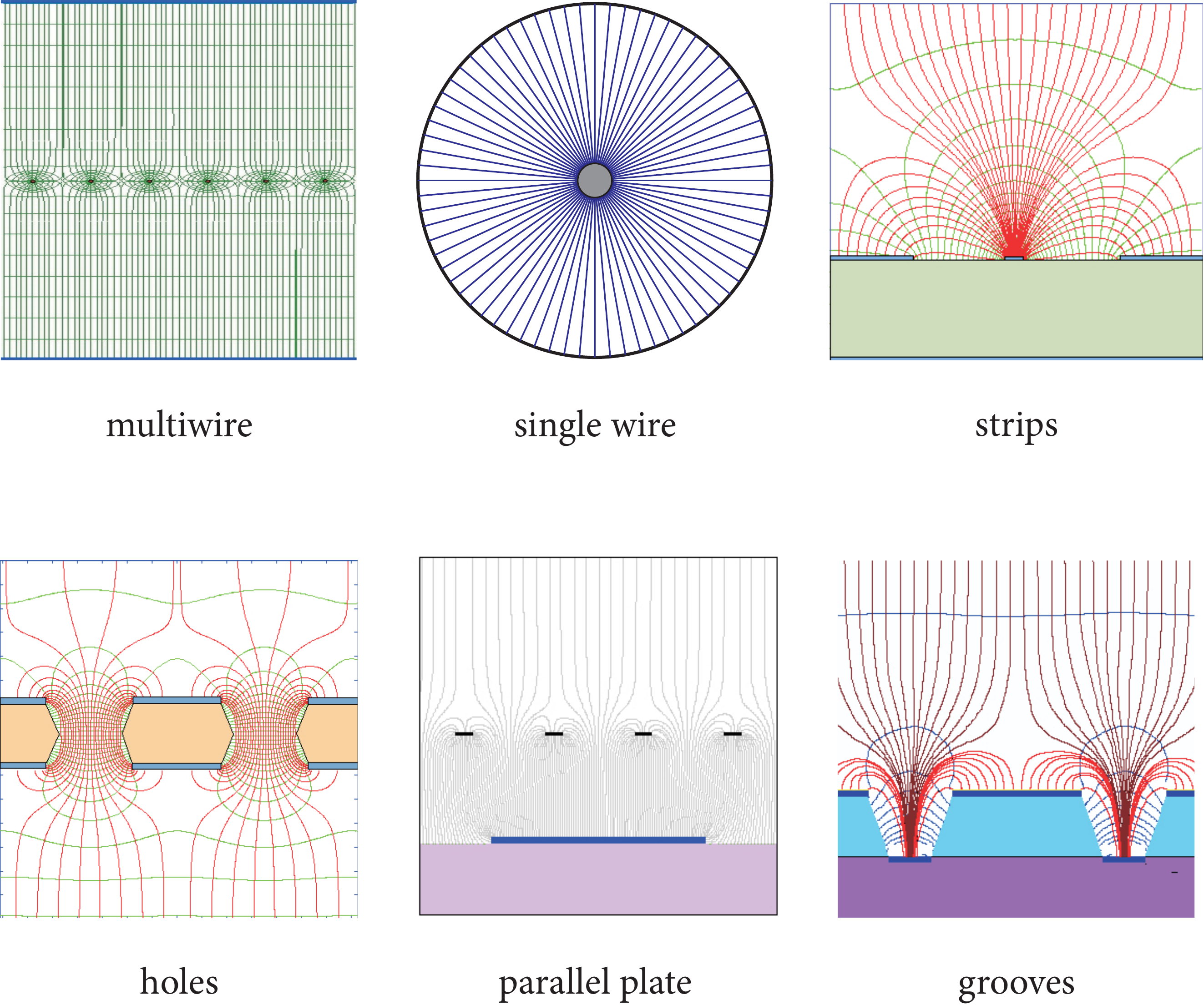}
\caption{Electric field patterns of various gas detector technologies. Each has a region of very strong field where multiplication takes place.}
\label{fieldmaps}
\end{figure}

\noindent
The working principle of all gas detectors is similar: radiation causes ionization in the gas, electrons and ions drift apart in an electric field, and the electrons create further electron-ion pairs in an avalanche process in a region with a strong electrostatic field.
Gaseous detectors differ in how this strong field region is created; many examples will be given to illustrate this.
For several decades the most popular way was using thin wires, either one or many, where close to the wire the field strength is inversely proportional to the distance to the wire.
This is illustrated in Fig.~\ref{fieldmaps}, the first two pictures.
The avalanche takes place few tens of microns from the wire, and the electrons are collected immediately.
The ions drift back all the way to the cathode; the signals from proportional wires are therefore almost entirely based on the movement of ions.

In recent years, many planar structures have emerged that generate an enhanced field region in various ways.
Several examples are shown in Fig.~\ref{fieldmaps} and still many more have been developed.
Common feature among all these structures is a narrow amplification gap of typically 50--100 microns, compared to many millimeters for wire-based structures.
These devices are now known under the common name of micropattern gaseous detectors (\textsc{mpgd}s).

\subsection{Microstrip gas chamber}
\begin{figure}[b]
\center\includegraphics[width=\columnwidth]{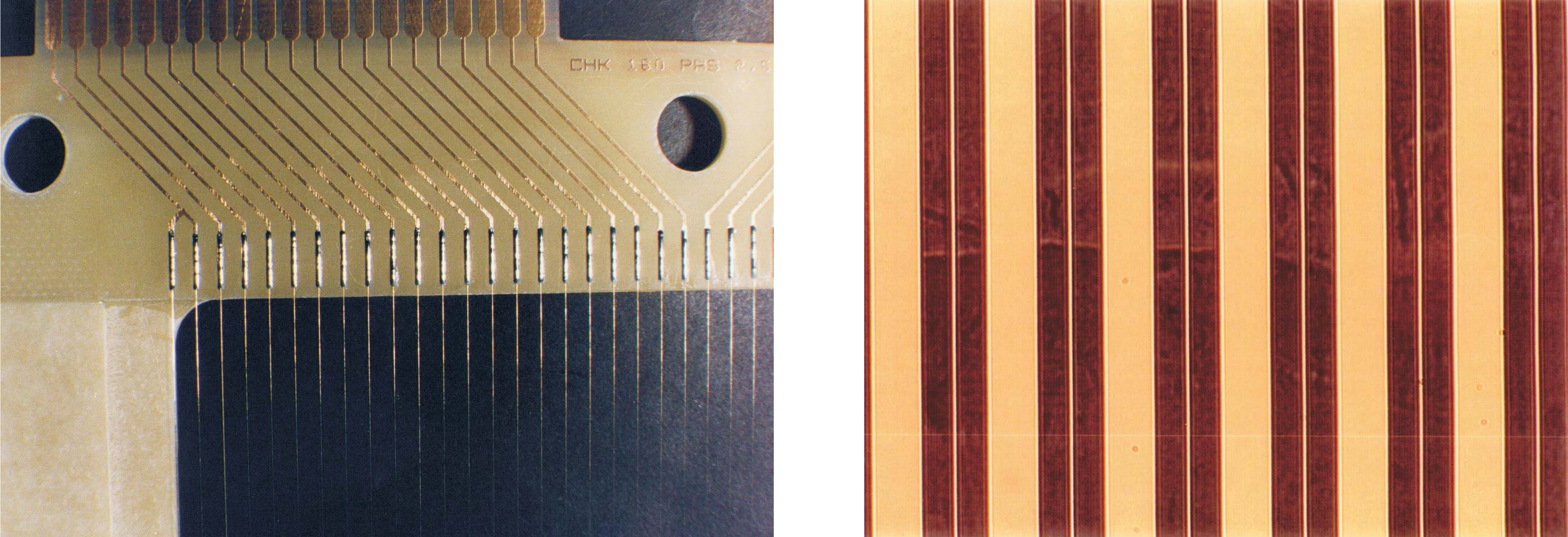}
\caption{Left: wires of a multiwire proportional chamber (\textsc{mwpc}) soldered to a frame. Right: microscope image of a microstrip gas chamber (\textsc{msgc})}
\label{mwpc&msgc}
\end{figure}
\begin{figure}
\center\includegraphics[width=\columnwidth]{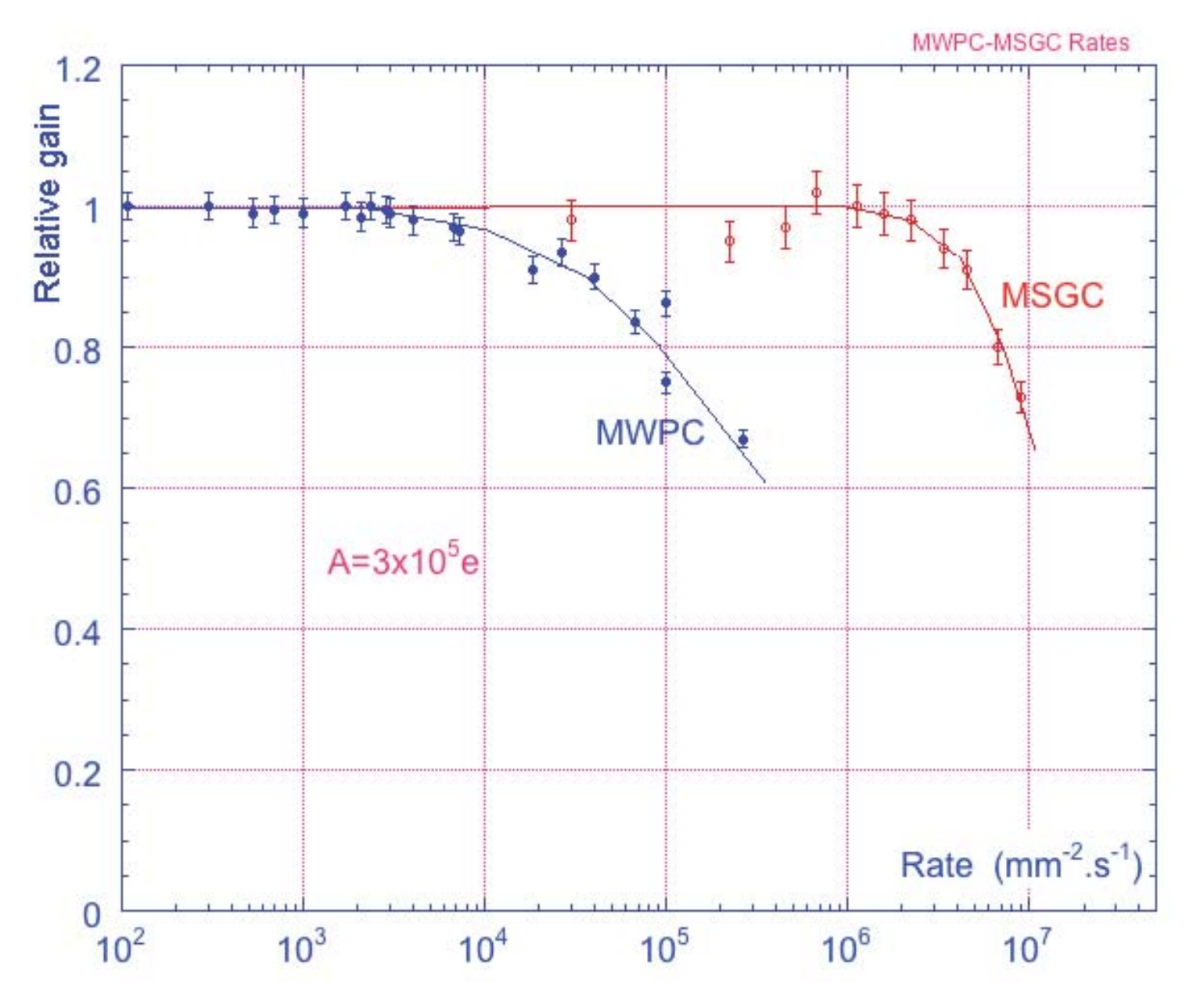}
\caption{Gain as a function of particle rate in otherwise constant conditions, for wire chambers in blue and \textsc{msgc}s in red.}
\label{ratecapability}
\end{figure}

\noindent The first such structure to gain popularity was the microstrip gas chamber~\cite{FirstMSGC} (\textsc{msgc}), of which the field pattern is shown in the third picture in Fig.~\ref{fieldmaps}.
The principle of an \textsc{msgc} resembles a wire chamber, with fine printed strips instead of thin wires, see Fig.~\ref{mwpc&msgc}.
Due to the microelectronics techniques employed in manufacturing the spacing between anode strips was as narrow as 200 microns, compared to at least several millimeters for wire chambers.
Most ions created in the avalanche process drift to the wider cathode strips, which are spaced only 60 microns away from the anodes.
This short drift path for ions overcomes the space charge effect present in wire chambers, where the slowly drifting ions may remain in the gas volume for milliseconds, and modify the electric field (thereby reducing the gain).
Figure~\ref{ratecapability} shows how this space charge effect limits the rate capability of wire chambers, and how the fine granularity of \textsc{msgc}s pushes this limit by two orders of magnitude.

The high rate capability of the \textsc{msgc} made it an attractive technology for many applications.
However, the development of the \textsc{msgc} also indicated some new limitations, most of which are common to all micropattern devices.
One common issue is the charging of insulating surfaces which modifies the field shape locally, limiting the time stability.
For \textsc{msgc}s this could be solved by surface treatment of the glass substrate to decrease the surface resistivity.

\begin{figure}[b]
\center\includegraphics[width=\columnwidth]{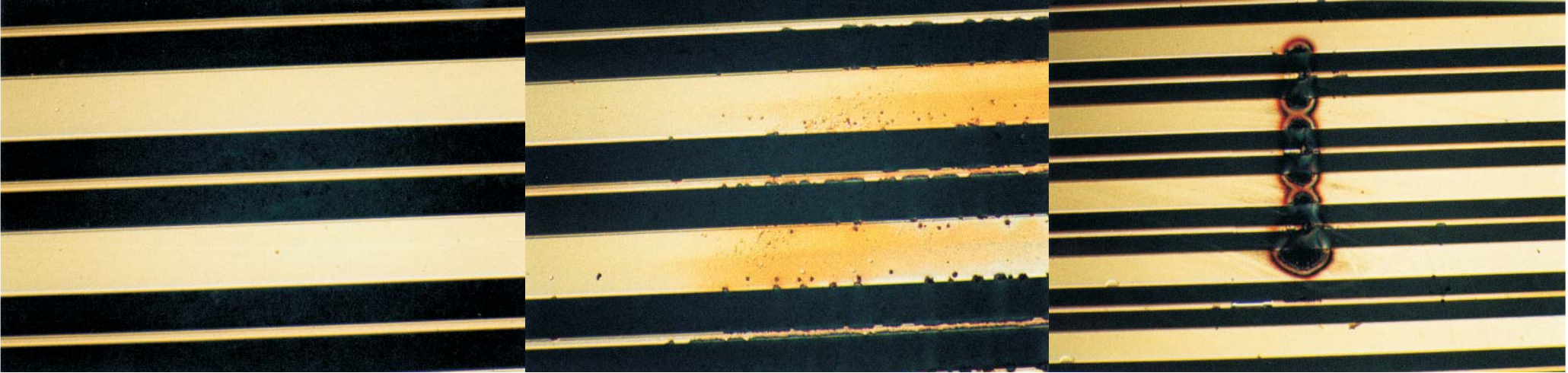}
\caption{Damage done to \textsc{msgc}s by discharges. In the rightmost frame anode strips are cut, leaving part of those anodes inactive. With its very thin metal layers \textsc{msgc}s are particularly vulnerable for discharge damage.}
\label{discharge}
\end{figure}
Possibly most important is the issue of discharges, which eventually led high-energy physics experiments to abandon \textsc{msgc} technology.
\textsc{Msgc}s suffered severely from such discharges, induced by heavily ionizing particles or high particle rates, which could fatally damage the fragile anode strips, see figure~\ref{discharge}.
In 1997 the gas electron multiplier (\textsc{gem}) was introduced~\cite{FirstGEM} as a preamplification stage for the \textsc{msgc}.
This allowed the \textsc{msgc} to work at a lower voltage, thereby lowering the probability of discharges as well as the energy involved in discharges when they occurred.
The \textsc{gem} principle was so successful that it soon became the basis for a detector in its own right.

\subsection{Gas electron multiplier}
\begin{figure}[t]
\center\includegraphics[width=\columnwidth]{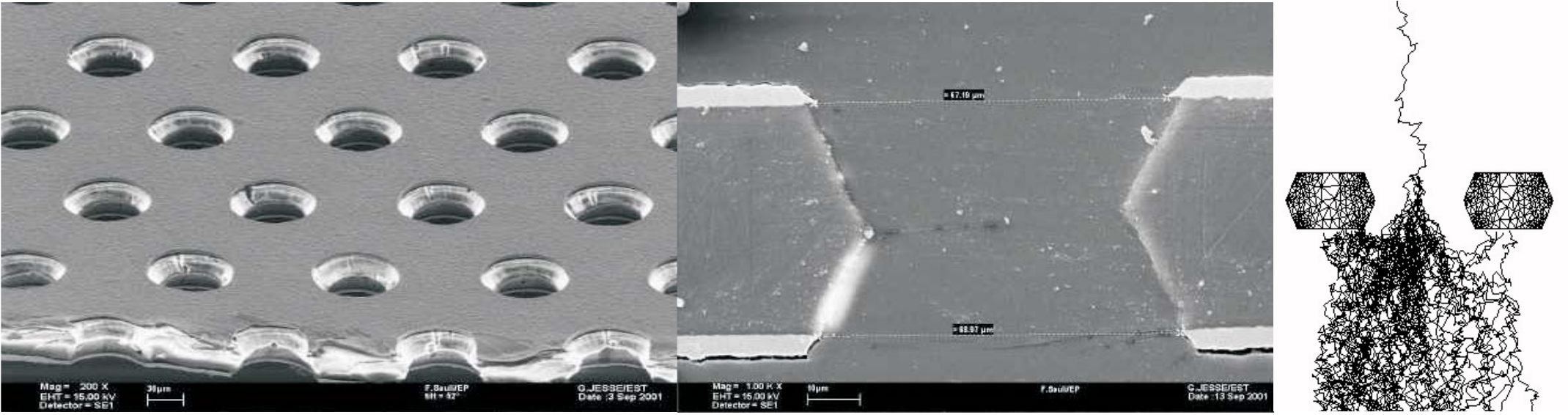}
\caption{Electron microscope images of a \textsc{gem} foil, and a simulated electron avalanche in a \textsc{gem} hole.}
\label{GEM}
\end{figure}
\begin{figure}[b]
\center\includegraphics[width=\columnwidth]{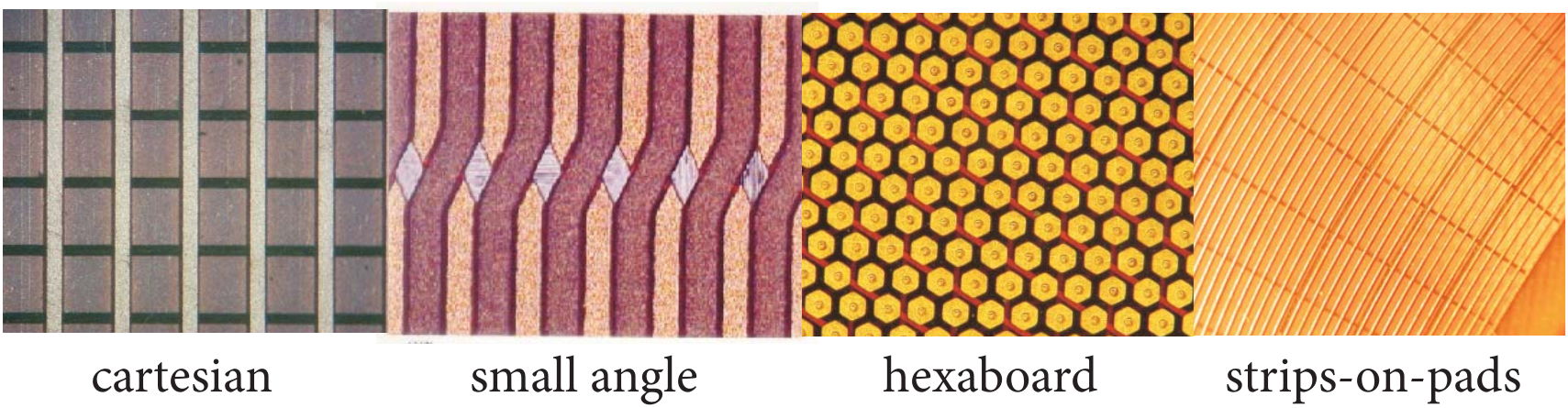}
\caption{Some examples of readout structures developed for \textsc{gem} detectors.}
\label{readouts}
\end{figure}
\begin{figure}
\center\includegraphics[width=\columnwidth]{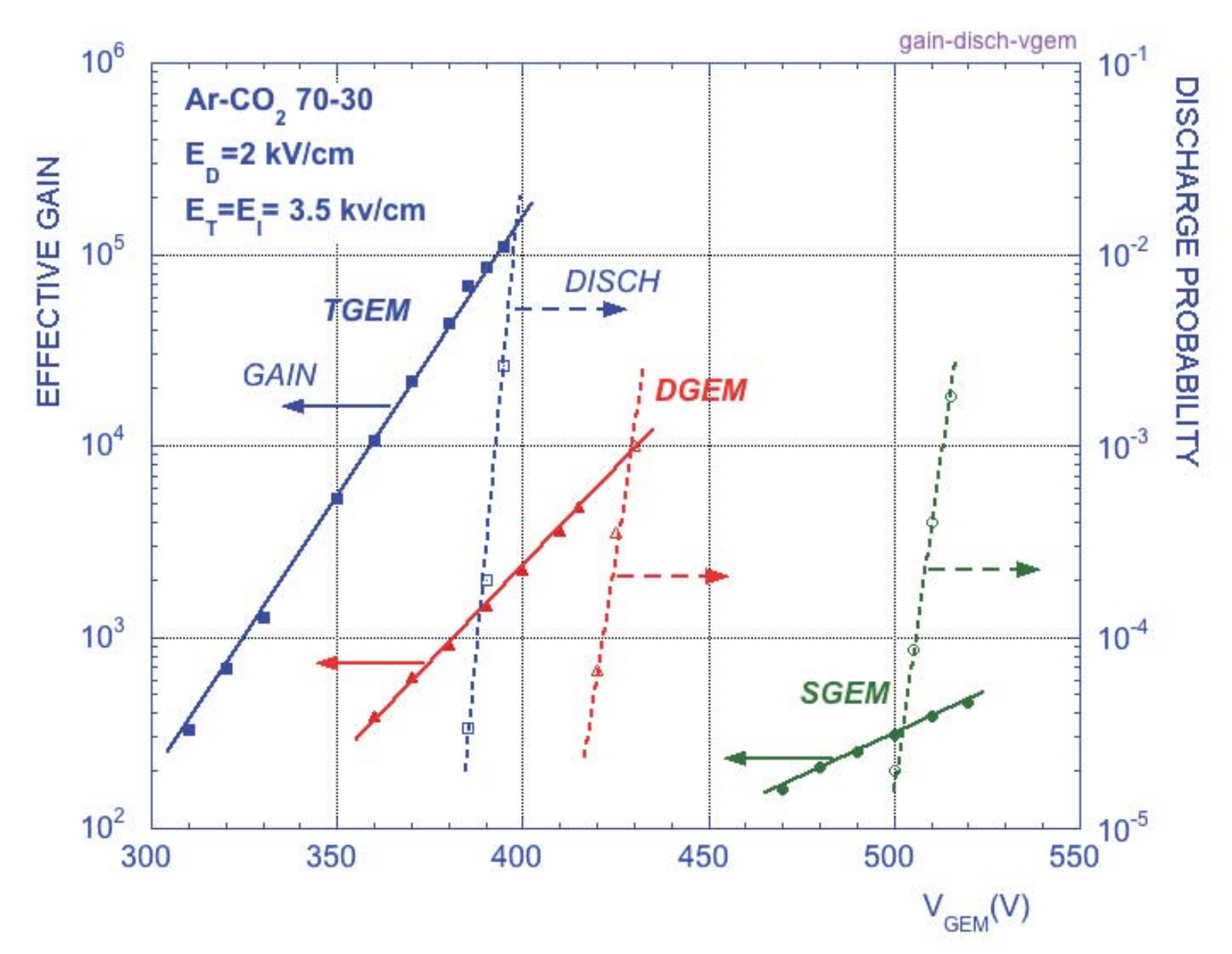}
\caption{Gain and discharge probability as a function of \textsc{gem} voltage, for single, double and triple \textsc{gem} detectors. Discharge probability is measured by irradiation with $\alpha$-particles, which are so strongly ionizing that they are likely to cause a discharge.}
\label{DischargeProb}
\end{figure}

\noindent The gas electron multiplier is a copper clad polyimide foil with a regular pattern of densely spaced holes, see Fig.~\ref{GEM}.
Upon applying a voltage between top and bottom electrodes, a dipole field is formed which focuses inside the holes where it is strong enough for gas amplification.
As a \textsc{gem} is only an amplification structure it is independent of the readout structure, which can be optimized for the application (see a few examples in Fig.~\ref{readouts}).
Due to the separation from the readout structure, possible discharges do not directly impact the front-end electronics, thus making the detector more discharge tolerant.
Also, it can be cascaded to achieve higher gain at lower \textsc{gem} voltage, which decreases the discharge probability, see Fig.~\ref{DischargeProb}.
The triple \textsc{gem} has now become a standard which is used in many high rate applications~\cite{COMPASS, TOTEM, LHCb}.

\subsection{Micromegas}
\begin{figure}[b]
\center\includegraphics[width=\columnwidth]{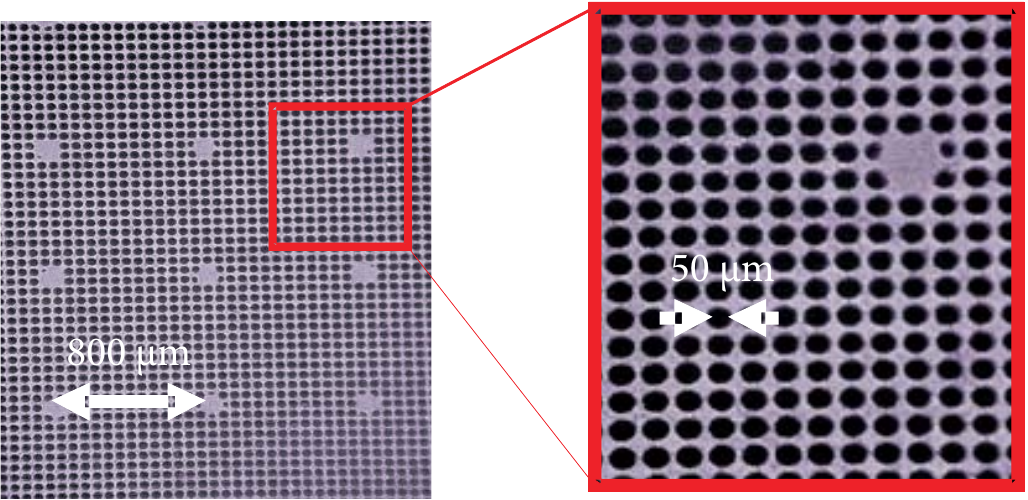}
\caption{Microscope images of a Micromegas detector, with indicated mesh and pillar spacings.}
\label{Micromegas}
\end{figure}

\noindent Another detector structure developed about the same time is the micromesh gas detector, or Micromegas~\cite{Micromegas}.
This detector has a parallel plate geometry with the amplification gap between a micromesh and the readout board.
Parallel plate amplification existed before, but the Micromegas has a much narrower amplification gap of around 50--100 $\mu$m.
The narrow amplification gap provides fast signals and a high rate capability.
The micromesh is supported by regularly spaced pillars which maintain the accurate spacing.
This is shown in Fig.~\ref{Micromegas}.

\section{Current trends in MPGDs}
\noindent The development of \textsc{mpgd}s took off in the 1990s mainly as a way to achieve a higher rate capability with gaseous detectors.
Since then applications have driven developers to exploit the additional benefits of these structures, such as excellent time and position resolution, resistance to aging, and intrinsic ion and photon feedback suppression.
Advances in available techniques for microelectronics and printed circuits opened ways to make new structures and optimize existing ones.
This led to a wide range of detector structures for an even wider range of applications, with a performance superior to any traditional gas detector.

\subsection{Techniques}
\noindent The techniques that enabled the advent of micropattern gas detectors come from the industry of microelectronics and printed circuits.
The microstrip gas chamber was made by employing photolithographic techniques used by microelectronics manufacturers.
Instead of silicon wafers, thin glass plates were used as a substrate for printing the fine strip patterns.
These glass plates were doped or sputter coated with so-called Pestov glass in order to reduce slightly the surface resistivity, which improved the time stability~\cite{SurfaceResistivity}.

The very thin metal layers of \textsc{msgc}s (few hundred nanometers) makes them vulnerable for discharges, which can easily do fatal damage (Fig.~\ref{discharge}).
Many of the later micropattern devices use thicker metals (few microns), and performance is normally unaffected by thousands of discharges.
The techniques used to pattern these metals and the insulators separating them come from the manufacturing of printed circuit boards (\textsc{pcb}s).
An advantage is the much lower cost, and the possibility to cover large areas.
These techniques include photolithography, metal etching and screen printing.

A rather special technique, thoroughly refined in the \textsc{cern pcb} workshop, is the etching of polyimide.
This is the basis of a number of micropattern gas detectors, including the \textsc{gem}.
Another more industry standard method to pattern insulators is using photo-imageable polymers, such as photoresist, coverlayers and solder masks.

Resistive layers or patterns are frequently used in \textsc{mpgd}s to quench discharges or to control the spreading of signals over readout channels when the anode is capacitively coupled to the readout elements.
Techniques to apply these layers range from laminating sheets of carbon-loaded polymers and spraying (no pattern) to screen printing and filling an insulating pattern with resistive paste.

\subsection{Technologies}
\begin{figure*}
\center\includegraphics[width=\textwidth]{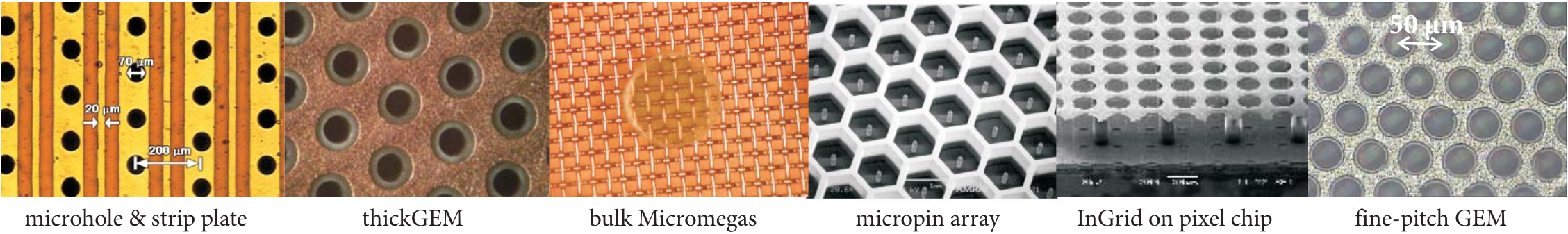}
\caption{Microscope images of various detector structures. See text for details on each frame.}
\label{technologies}
\end{figure*}
\noindent 
A few of the most prominent micropattern gas detector technologies have been mentioned in the introduction.
Many more types of structures were developed and are currently used, which are often derived from \textsc{msgc, gem} or Micromegas.
A few more examples are discussed here, but the selection is by no means exhaustive.

The refinement of the polyimide etching technique that is used to make \textsc{gem}s, is also used for some detectors with a readout structure in the same plane as the amplification structure.
These are the \emph{\textsc{well}}~\cite{WELL} and the \emph{groove detector}~\cite{groove}.
Unlike the \textsc{gem} these structures are not ``transparent'', all the electrons from the avalanche are collected on the bottom electrode which is also the readout structure.
The \emph{microhole and strip plate}~\cite{MHSP} combines the amplification mechanisms of \textsc{gem} holes and microstrips (see Fig.~\ref{technologies}, first frame), and combines a high gas gain with an unparalleled ion feedback suppression.

Another \textsc{gem}-derivative is the \emph{thick\textsc{gem}}~\cite{thickGEM}, also shown in Fig.~\ref{technologies}.
This is a hole-type amplification structure, where the flexible polyimide substrate is replaced by a thicker glassfiber-reinforced-epoxy plate and the holes are mechanically drilled.
The substrate is the standard base material for rigid \textsc{pcb}s and is therefore cheap, and readily available from any \textsc{pcb} manufacturer.
Also the automatic drilling of the holes is a standard industry procedure.
One has full control over the hole pitch and diameter, and the shape, size and thickness of the base material.
These structures are convenient for applications where position and time resolution are not the most critical parameters, but which require a high gain and a certain ruggedness.
Thick\textsc{gem}s are for instance popular for photodetector applications, where the stiff substrate lends itself well to the vacuum deposition of a CsI photoconverter~\cite{CsI}.
More recently, electrodes of thick\textsc{gem}s have been covered with or replaced by resistive layers~\cite{RetGEM}.
These detectors are reported to work stably in streamer mode, due to the enhanced quenching by the resistive layers.

Micromegas detectors underwent a technical improvement with the introduction of a new fabrication method~\cite{bulk}.
Here a woven metal micromesh is laminated to the readout board between layers of photoimageable soldermask.
These soldermask layers can subsequently be patterned by \textsc{uv}-exposure to create the supporting pillar structure (see the third frame of Fig.~\ref{technologies}).
The materials involved are quite inexpensive, and the processes are industry standard, which makes it suitable for large scale production.
Also, the homogeneity of the grid spacing is better than of the original Micromegas detectors, and the detector is very robust.

The \emph{micropin array}~\cite{MicroPin} was introduced for x-ray imaging (see Fig.~\ref{technologies}, fourth frame).
The spherical geometry of the electric field close to the end of each pin (proportional to $1/r^2$ compared to $1/r$ of a wire chamber) gives rise to very short amplification region, allowing a rate-stable high gain.
A similar philosophy led to the development of the \emph{microdot chamber}~\cite{MicroDot}, for which microelectronics techniques were employed to reach feature sizes of only a few microns.

The coming of age of \emph{post-wafer processing} techniques marked the introduction of \textsc{mpgd}s with pixel readout.
These detectors use the bump-bonding pads of a pixel chip as a readout structure.
The position and time resolution of these devices is unmatched by any other gas detector.
Due to their high sensitivity they can distinguish each primary electron.
This enables them to resolve delta-rays from a track or to reconstruct the direction of emission of a photoelectron from an x-ray conversion (related to the x-ray polarization).
One group uses a Micromegas-type of gas amplification: \emph{InGrid}~\cite{GridPix}.
The grid electrode and the insulating pillar structure supporting it are made directly on the chip by post-wafer processing techniques, allowing the grid holes to be aligned with the readout pads (see Fig.~\ref{technologies}, fifth frame).
Another group uses an \textsc{asic} with a hexagonal readout pad structure, and a \textsc{gem}-based amplification structure~\cite{GEMpix}.
Here the \textsc{gem} has a reduced pitch of $50 \mu$m and thickness of $25 \mu$m (compared to $140 \mu$m  and $50 \mu$m respectively for standard \textsc{gem}s) to match the granularity of the readout (see the last frame of Fig.~\ref{technologies}).

\subsection{Applications}
\noindent Micropattern gas detectors have already been applied in many instruments and experiments, both by science and industry.
Possible fields of application are high-energy and nuclear physics, synchrotron and thermal neutron research, medical imaging and homeland security.
Most structures were primarily developed for high rate tracking of charged particles in nuclear and high-energy physics experiments.
For instance Micromegas~\cite{COMPASSMM} and \textsc{gem}s~\cite{COMPASS} are used in the \textsc{compass} experiment, and \textsc{gem}s in \textsc{lhc}b~\cite{LHCb} and \textsc{totem}~\cite{TOTEM} experiments.
Also for the \textsc{lhc} machine upgrade program to increase its luminosity by roughly a factor of ten, most of the experiments foresee replacement of wire chambers, drift tubes and resistive plate chambers by \textsc{mpgd}s.
However many \textsc{mpgd}s have shown to be suitable for other applications as well.
A few examples are given here.

Both \textsc{gem}s and Micromegas can be used for the readout of a time projection chamber~\cite{TPC} (\textsc{tpc}).
Compared to wire chambers, these \textsc{mpgd}s have the benefit that the planar structure suppresses the so-called $\mathbf{E}\times\mathbf{B}$ effects which limit the spatial resolution of wire chambers in \textsc{tpc} configuration.
Also, both Micromegas and \textsc{gem}s have a natural ion feedback suppression, which may make a gating structure unnecessary.

As mentioned before, \textsc{gem}-like structures can be coated with a photoconverter (typically CsI) to serve as a photon counter.
In this way, large areas can be covered with hardly any dead zones, and the technique is inexpensive.
This makes it attractive for ring imaging Cherenkov detectors, of which the photodetector planes often span several square meters.
Also here the ion feedback suppression is an added benefit, as it increases the lifetime of the photoconverter.
In addition, the detector can be made ``hadron-blind'' by reversing the drift field, and even ``windowless'' if the Cherenkov radiator gas (in that case typically CF$_4$) is also used as amplification gas~\cite{HBD}.

X-ray counting and imaging detectors can be based on \textsc{mpgd}s \cite{x-ray}, as x-rays convert in some noble gases leaving typically few hundred primary electrons for detection.
For these purposes efficient x-ray conversion gases are frequently used, such as xenon or krypton.
Argon is about an order of magnitude less efficient, but so much cheaper that it can still be attractive for high rate applications.

Microstrip gas chambers and \textsc{gem} detectors are used as neutron detectors~\cite{boron}.
Typically a boron layer (in the form of B$_2$O$_3$) is evaporated onto the \textsc{gem} foils, which acts as a neutron converter via the reaction $^{10}\text{B} + \text{n} \rightarrow\  ^7\text{Li} + \alpha$.
In the case of \textsc{msgc}s, $^3$He is often used as both amplification and convertor gas.
Here the conversion reaction is: $^3\text{He} + \text{n} \rightarrow\  ^3\text{H} + \text{p}$.

\subsection{Performance}
\noindent Depending on the application, the performance of \textsc{mpgd}s has different figures of merit.
The first \textsc{mpgd}s were designed to obtain a high rate capability.
Several MHz/mm$^2$ of charged particles are easily reached with, for instance, a triple \textsc{gem} detector, without a measurable loss of gain and with negligible discharge probability.

Time, position and energy resolution are crucial figures for most applications.
\textsc{Gem}-based detectors normally have a position resolution of about $50 \mu$m, Micromegas can go down to $\sim 12 \mu$m if equipped with a high density readout board.
Time resolutions are of the order of few nanoseconds.
X-ray energy resolution is often measured using a $^{55}$Fe source, obtaining a \textsc{fwhm} between 15\% and 22\%.
\textsc{Mpgd}s with pixel chip readout report position resolutions below $10 \mu$m and a time resolution of 1 ns.
From the $^{55}$Fe spectrum they can resolve the K$_\alpha$ and K$_\beta$ energies, and reach a resolution of 12\%.

The reduction of ion backflow into the drift region is a general property of \textsc{mpgd}s.
It is usually expressed as a fraction of the effective gain, and this value depends quite strongly on the way the fields are configured in the chamber.
Microhole and strip plates feature a particularly effective ion feedback suppression of the order $10^{-4}$ in optimized conditions.

Aging modes of gas detectors are largely understood in the case of wire chambers.
There the plasmas that are formed during avalanches in the strong field near the wire deposit layers of silica or polymers which reduce the gain and give rise to micro discharges.
Most micropattern devices do not generate such a strong field at the surface of the conductors, and consequently little signs of aging have been observed.
Aging studies of \textsc{mpgd}s specifically have rarely been done yet, and time will prove if they are as resistant to aging as it seems.

\section{An r\textit{\&}d collaboration for MPGDs}
\noindent RD51 is a large \textsc{r\&d} collaboration, which unites many institutes in an effort to advance technological development of micropattern gas detectors.
At the time of writing there are $\sim 430$ participating authors from 75 institutes in 25 countries worldwide.
The efforts of the collaboration do not focus on one or a few particular applications for \textsc{mpgd}s, but is rather \emph{technology oriented}.
It is a platform for sharing of information, results and experiences, and for steering \textsc{r\&d} efforts.
It tries to optimize the cost of \textsc{r\&d} projects by sharing resources, creating common projects and providing common infrastructure.

\subsection{Organization}

\begin{table*}
\caption{Organization of RD51 in working groups and tasks.}
\label{WGs}
\begin{tabular}{@{}l@{}c@{}c@{}c@{}c@{}c@{}c@{}c@{}}
\toprule
&\textsc{Wg1}&\textsc{Wg2}&\textsc{Wg3}&\textsc{Wg4}&\textsc{Wg5}&\textsc{Wg6}&\textsc{Wg7}\\
&\textsc{Mpgd} technology&Characterization&Applications&Software&Electronics&Production&Common test\\
&\& new structures&\& physics issues&&\& simulation&&&facilities\\
\midrule
\multirow{8}{5mm}{\begin{sideways}\parbox{5mm}{\textsc{Objectives}}\end{sideways}}&
\multirow{-2}{26mm}{\center \vspace{-1mm}Design\\optimization.\\Development of\\new geometries\\and techniques}&
\multirow{-2}{26mm}{\center \vspace{-1mm}Common test\\standards.\\Characterization\\of physical pheno-\\mena in \textsc{mpgd}s}&
\multirow{-2}{25mm}{\center \vspace{-1mm}Evaluation and\\optimization\\for specific\\applications}&
\multirow{-2}{24mm}{\center \vspace{-1mm}Development\\of common\\software and\\documentation\\for \textsc{mpgd}s}&
\multirow{-2}{26mm}{\center \vspace{-1mm}Readout\\electronics\\optimization and\\intergration with\\\textsc{mpgd}s}&
\multirow{-2}{25mm}{\center \vspace{-1mm}Development\\of cost-effective\\technologies and\\industrialization}&
\multirow{-2}{23mm}{\center \vspace{-1mm}Sharing of\\common\\infrastructure\\for detector\\characterization}\\[15mm]
\midrule
\multirow{15}{5mm}{\begin{sideways}\parbox{5mm}{\textsc{Tasks}}\end{sideways}}&
\multirow{-2}{26mm}{\center \vspace{-1mm}Large area\\\textsc{mpgd}s\\[-.5mm]---\\[-.5mm]Design\\optimization\\New geometries\\Fabrication\\[-.5mm]---\\[-.5mm]Development\\of rad-hard\\detectors\\[-.5mm]---\\[-.5mm]Development\\of portable\\detectors}&
\multirow{-2}{26mm}{\center \vspace{-1mm}Common test\\standards\\[-.7mm]---\\[-.7mm]Discharge\\protection\\[-.7mm]---\\[-.7mm]Aging and\\radiation\\hardness\\[-.7mm]---\\[-.7mm]Charging-up\\and rate\\capability\\[-.7mm]---\\[-.7mm]Avalanche\\statistics}&
\multirow{-2}{25mm}{\center \vspace{-1mm}Tracking and\\triggering\\[-1mm]---\\[-1mm]Photodetection\\[-1mm]---\\[-1mm]Calorimetry\\[-1mm]---\\[-1mm]Cryogenic det.\\[-1mm]---\\[-1mm]X-ray \& neutron\\imaging\\[-1mm]---\\[-1mm]Astroparticle\\physics appl.\\[-1mm]---\\[-1mm]Medical appl.\\[-1mm]---\\[-1mm]Plasma diagn.\\Homeland sec.}&
\multirow{-2}{24mm}{\center \vspace{-1mm}Algorithms\\[1mm]---\\[1mm]Simulation\\improvements\\[1mm]---\\[1mm]Common\\platforms\\(\textsc{root}, Geant4)\\[1mm]---\\[1mm]Electronics\\modeling}&
\multirow{-2}{26mm}{\center \vspace{-1mm}\textsc{Fe} electronics\\requirements\\definition\\[-1mm]---\\[-1mm]General purpose\\pixel chip\\[-1mm]---\\[-1mm]Large area\\systems with\\pixel readout\\[-1mm]---\\[-1mm]Portable multi-\\channel system\\[-1mm]---\\[-1mm]Discharge\\protection\\strategies}&
\multirow{-2}{25mm}{\center \vspace{-1mm}Common\\production\\facility\\[1mm]---\\[1mm]Industrialization\\[1mm]---\\[1mm]Collaboration\\with\\industrial\\partners}&
\multirow{-2}{23mm}{\center \vspace{-1mm}Testbeam\\facility\\[1mm]---\\[1mm]Irradiation\\facility}\\[45mm]
\bottomrule
\end{tabular}
\end{table*}

\noindent 
RD51 has two spokespersons, Leszek Ropelewski\footnote{\href{mailto:Leszek.Ropelewski@cern.ch}{\texttt{Leszek.Ropelewski@cern.ch}}} and Maxim Titov\footnote{\href{mailto:maxim.titov@cea.fr}{\texttt{maxim.titov@cea.fr}}}, who can be contacted for more information.
Concerning all scientific matters the collaboration is governed by a \emph{collaboration board} (\textsc{cb}), which is also responsible for coordinating the financial planning and other resource issues, in particular for managing the common fund.
Representatives from all collaborating institutes are seated in the \textsc{cb}, and have voting rights.
A \emph{management board} (\textsc{mb}) supervises the progress of the work program along the lines defined by the \textsc{cb} and prepares decisions for and makes recommendations to the \textsc{cb}.

The activity is divided in seven working groups (\textsc{wg}s), covering all relevant topics of \textsc{mpgd}-related \textsc{r\&d}.
A number of tasks is assigned to each working group.
Table~\ref{WGs} lists all the \textsc{wg}s and indicates their objectives and tasks.

\textsc{Wg}1 is concerned with the technology of \textsc{mpgd}s and the design of new structures.
Examples are efforts to make Micromegas, \textsc{gem} and thick\textsc{gem} technologies suitable for large areas~\cite{LargeGEM}.
Also interesting is the development of cylindrical \textsc{gem}~\cite{cylindricalGEM} and Micromegas~\cite{cylindricalMM} detectors for inner barrel tracking.
A recent development is the introduction of spherical \textsc{gem}s~\cite{spherical} for parallax-free x-ray diffraction measurements.

The second working group deals with physics issues of \textsc{mpgd}s, such as discharges, charging of dielectric surfaces and aging.
Also, common test standards are proposed to enable different groups to compare their results.
Regular meetings have become a forum for exchanging results and for discussion about what are actually the most fundamental properties of micropattern gas detectors.

\textsc{Wg3} concentrates on the applications of \textsc{mpgd}s, and on how to optimize detectors for particularly demanding applications.
Examples have been listed above and new applications still appear.
Sometimes from surprising fields: one project aims to construct very large area \textsc{gem} chambers to detect nuclear fission materials or waste in cargo containers by tomography of cosmic ray muons~\cite{tomography}.

\textsc{Wg4} develops simulation software and makes progress in the field of simulation.
Simulation is essential to understand the behavior of detectors.
A mature range of software tools is available for simulating primary ionization (Heed\footnote{Author: Igor Smirnov (\href{http://consult.cern.ch/writeup/heed/}{\texttt{\scriptsize http://consult.cern.ch/writeup/heed/}})}), electron transport properties in gas mixtures in electric and magnetic fields (Magboltz\footnote{Author: Stephen Biagi (\href{http://consult.cern.ch/writeup/magboltz/}{\texttt{\scriptsize http://consult.cern.ch/writeup/magboltz/})}}), and gas avalanches and induction of signals on readout electrodes (Garfield\footnote{Author: Rob Veenhof (\href{http://garfield.web.cern.ch/garfield/}{\texttt{\scriptsize http://garfield.web.cern.ch/garfield/}})}).
Garfield has interfaces to Heed and Magboltz and only needs to be supplied with a field map and detector configuration.
A field map can be generated by commercial finite-element method (\textsc{fem}) programs such as Ansys, Maxwell, Tosca, QuickField and \textsc{Femlab}.
Within the collaboration, an open-source field solver is developed and recently released called ne\textsc{bem}~\cite{nebem}.
It is based on the boundary element method (\textsc{bem}), and is in most respects superior to \textsc{fem} solvers for gas detector simulations.
A recent development is the simulation of the \emph{Penning transfer} mechanism (an excited molecule of one gas component can cause ionization of a molecule of another component) in the modeling of ionization and avalanche processes, greatly improving the estimations of gas gain.

Front-end electronics and data acquisition systems are discussed in \textsc{wg5}.
Spearhead of this working group is the development of a scalable modular readout and acquisition system that can be applied to a single detector or to a large systems, and that can work with many different front-end \textsc{asic}s.
This system was recently commissioned, and is now being produced at larger scale.

\textsc{Wg6} deals with the production of \textsc{mpgd}s.
Almost all \textsc{mpgd}s were first made in the \textsc{cern pcb} workshop of Rui de Oliveira, and it remains an almost exclusive manufacturing site for most technologies.
Hence, efforts in \textsc{wg6} are aimed at plans for upgrading this workshop on one hand, and industrial partnership and export of the technology and know-how on the other.
Also, scenarios are developed for industrial scale production of some \textsc{mpgd}s (especially \textsc{gem}s and Micromegas), in case a large experiment decides to implement them in their system.

Finally, \textsc{wg7} coordinates the effort to set up a shared test infrastructure in the form of test beam and irradiation facilities.
The test beam facility is equipped with supply and exhaust of gases, including flammable mixtures.
Also a large 1.4 Tesla magnet is provided.
The irradiation facility will provide a strong gamma source (a 10 TBq $^{137}$Cs source is foreseen) combined with a 100 GeV muon test beam ($10^4$ muons per spill) and is called \textsc{gif++}~\cite{GIF++}.

\section{Conclusions and contacts}
\noindent Micropattern gas detectors have a great potential in science and industry, in medical and commercial applications.
RD51 is committed to fulfill this potential.
The collaboration welcomes new institutes who are interested in participating in the development of micropattern gas detectors.
Up-to-date information and relevant contacts can be found on the collaboration webpage\footnote{\href{http://rd51-public.web.cern.ch/RD51-Public/}{\texttt{http://rd51-public.web.cern.ch/RD51-Public/}}}, or by contacting the spokespersons.

\pagebreak
\end{document}